\documentclass{article}
\usepackage{graphicx}
\usepackage[english]{babel}

\title{
\includegraphics[width=0.35\textwidth]{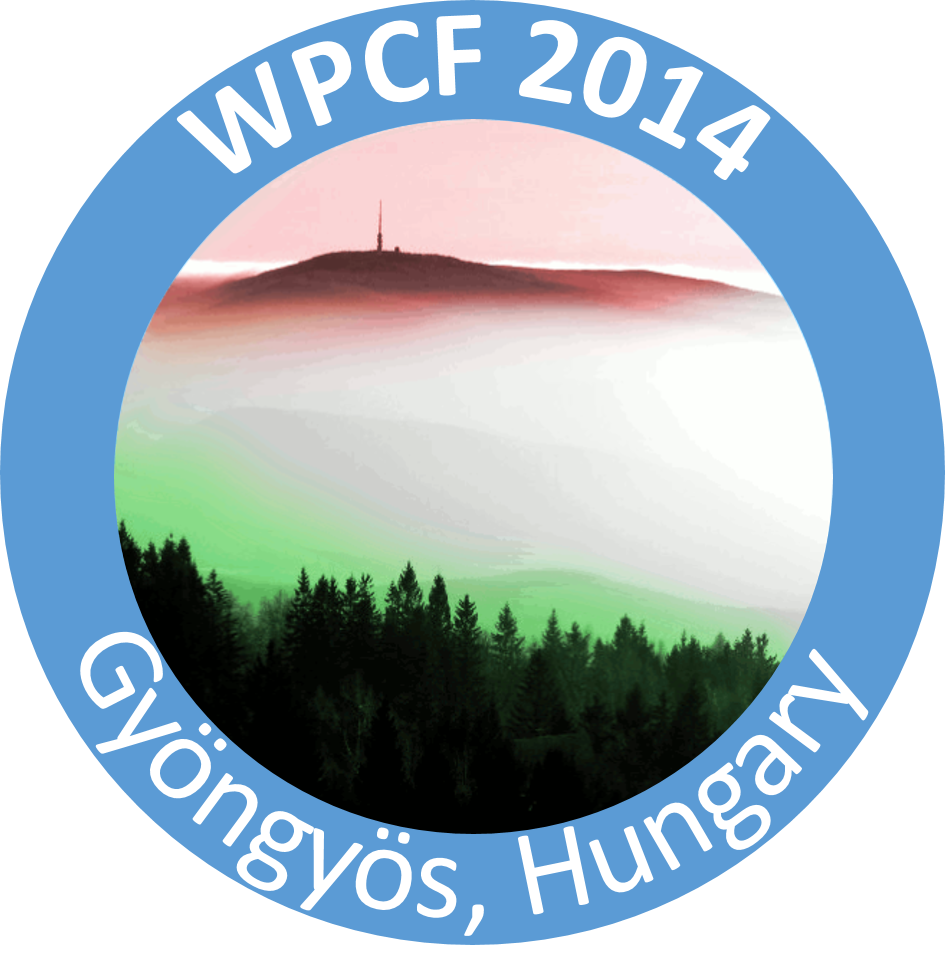}\\[1cm]
A Monte Carlo Study of Multiplicity Fluctuations in Pb-Pb Collisions at LHC Energies}
\author{{Ramni Gupta\footnote{email:ragupta@cern.ch}}\\[1ex]
Department of Physics \& Electronics, University of Jammu,  Jammu, India\\
}

\begin{document}

\fontfamily{lmss}\selectfont
\maketitle

\begin{abstract}
With large volumes of data available from LHC, it has become possible to study the multiplicity distributions for the various possible behaviours of the multiparticle production in collisions of relativistic heavy ion collisions, where a system of dense and hot partons has been created. In this context it is important and interesting as well to check how well the Monte Carlo generators can describe the properties or the behaviour of multiparticle production processes. One such possible behaviour is the self-similarity in the particle production, which can be studied with the intermittency studies and further with chaoticity/erraticity, in the heavy ion collisions. We analyse the behaviour of erraticity index in central Pb-Pb collisions at centre of mass energy of 2.76 TeV per nucleon using the AMPT monte carlo event generator, following the recent proposal by R.C. Hwa and C.B. Yang, concerning the local multiplicity fluctuation study as a signature of critical hadronization in heavy-ion collisions. We report the values of erraticity index for the two versions of the model with default settings and their dependence on the size of the phase space region. Results presented here may serve as a reference sample for the experimental data from heavy ion collisions at these energies.

\end{abstract}

\section{Introduction}
Dynamics of the initial processes, that is the distributions and the nature of interactions of quarks and gluons,  in the heavy ion collisions affect the final distribution of the particles produced~\cite{Sgavin:2012aa}. Of the various distributions, multiplicity distributions play fundamental role in extracting first hand information on the underlying particle production mechanism. If QGP is formed at these energies the QGP-hadron phase transition is expected to be accompanied by large local fluctuations in the number of produced particles in the regions of phase space~\cite{Adamovich:1988pf}. Thus the study of fluctuations in the multiplicity is an important tool to understand the dynamics of initial processes and consequently the processes of strong interactions, phase transition and also to understand correlations of QGP formation~\cite{VanHove:1984sf}.

\par
A comprehensive theoretical model which can explain and give answers to all the complexities of the physics involved at high energy and densities, as is created in the heavy ion collisions,  is still not available. A successful model focussed on one aspect of the problem may not say much about the other aspects, but at least should not contradict what is observed. The measures which are studied in the present work rely on the large bin multiplicities. At LHC energies multiplicities are high and it is possible to have detailed study of the local properties in ($\eta,\phi$) space for narrow $p_{T}$ bins and thus to explore the dynamical properties of the system created in the heavy ion collisions.  Thus as an initial attempt to understand the nature of global properties, as manifested in local fluctuations,  here we develop and test the methodology and   effectiveness of the analysis, analysing simulated events for Pb-Pb collisions at              $\sqrt{s_{NN}}$= 2.76 TeV using A Multi-Phase Transport (AMPT) model.  

\par
Study of charged particle multiplicity fluctuations is one of the sensitive probes to learn about the properties of the system produced in the heavy ion collisions. Factorial moments are one of the convenient tools for studying fluctuations in the particle production. The concept of factorial moments was first used by A.~Bialas and R.~Peschanski~\cite{Bialas:1988hf} to explain unexpectedly large local fluctuations in high multiplicity events recorded by the JACEE Collaboratin. Advantage of studying fluctuations using factorial moments is that these filter out statistical fluctuations. 
The normalised factorial moment $F_{q}$  is defined as

\begin{equation}
F_{q}(\delta^{d}) = \frac{\langle n!/(n-q)!\rangle}{\langle n \rangle^{q}}
\label{eqfact1}
\end{equation}

where $n$ is the number of particles in a bin of size $\delta^{d}$ in a $d$-dimensional space of observables and  $\langle \ldots \rangle$ is either vertical or horizontal averaging. $q$ is the order of the moment and is a positive integer $\geq 2$. Then a power-law behaviour 

\begin{equation}
F_{q}(\delta) \propto \delta^{-\varphi_{q}}
\label{inter1}
\end{equation}

over a range of small $\delta$ is referred to as  {\em intermittency}.   In terms of the number of bins $M \propto 1/\delta$, Eq.~\ref{inter1} may be written as
 
\begin{equation}
F_{q}(M) \propto M^{\varphi_{q}}
\label{inter2}
\end{equation}

where $\varphi_{q}$ is the intermittency index, a positive number. 

\par
Even if the scaling behaviour in Eq.~\ref{inter1} is not satisfied, to a high degree of accuracy, $F_{q}$ satisfies the power law behaviour
\begin{equation}
F_{q} \propto F_{2}^{\beta_{q}}
\label{fq}
\end{equation}

This is referred to as F-scaling. In attempts to quantify  systems undergoing second order phase transition, in Ginzburg-Landau (GL) theory~\cite{Hwa:1992df}, it is observed that

\begin{equation}
\beta_{q} = (q-1)^{\nu}, \qquad \nu=1.304,
\label{beta}
\end{equation}

 the scaling exponent, $\nu$, is essentially idependent of the details of the GL parameters.
 
\par
Factorial moments ($F_{q}$'s) do not fully account for the fluctuations that the system exhibits. Vertically averaged horizontal moments, can gauge the spatial fluctuations, neglecting the event space fluctuations. On the other hand, horizontally averaged vertical moments lose information about spatial fluctuations and only measure the fluctuations from event-to-event. {\em  Erraticity analysis} introduced in~\cite{Cao:1995cf}, where one finds moments of factorial moment distribution, takes into account the spatial as well as the event space fluctuations.  It measures  fluctuations of the spatial patterns and quantifies this in terms of an index named as {\em erraticity index} ($\mu_{q}$).   In a  recent work~\cite{Hwa:2012hy}, $\mu_{q}$ is observed to be a measure sensitive to the dynamics of the particle production mechanism and hence to the different classes of quark-hadron phase transition. 
\par
In erraticity analysis, event factorial moments are studied, defined for an  $e^{th}$ event  as  

\begin{equation}
F_{q}^{e}(M) = \frac{f_{q}^{e}(M)}{[f_{1}^{e}(M)]^{q}}
\label{eq1}
\end{equation}
  wherein,

\begin{equation}
f_{q}^{e}(M)   =  \langle n_{m}(n_{m}-1)......(n_{m}-q+1)\rangle_{h}
\label{eq2}
\end{equation}

\begin{figure}[t]
\centerline{\includegraphics[scale=0.38]{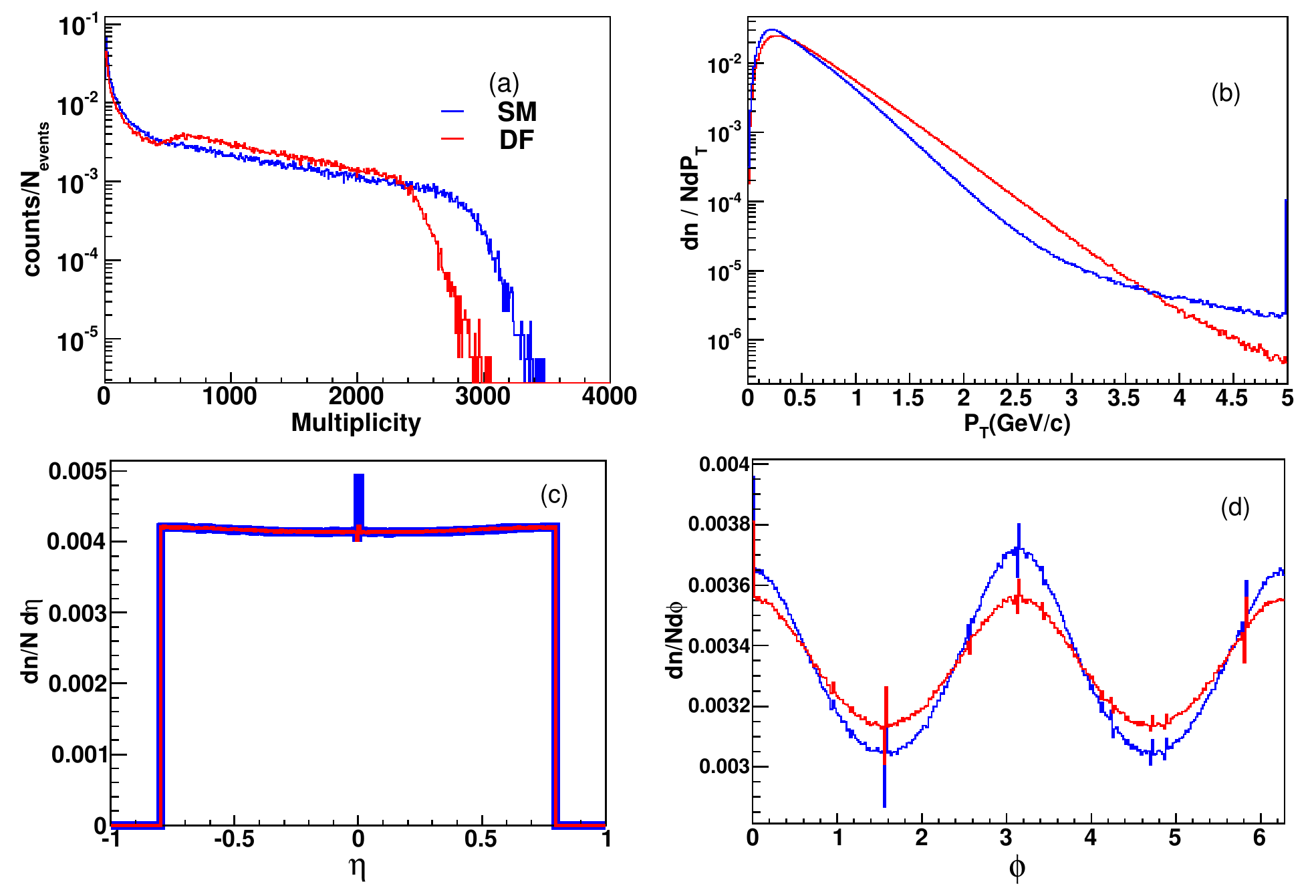}}
\caption{(a) Multiplicity distribution (b) $p_{T}$ (c) $\eta$ and (d) $\phi$  distributions of charged particles generated in Pb-Pb collisions at $\sqrt{s_{NN}}$ =2.76 TeV  using  the DF and  SM  AMPT.}
\label{fig1}
\end{figure}

\begin{figure}
\centering{\includegraphics[scale=0.3]{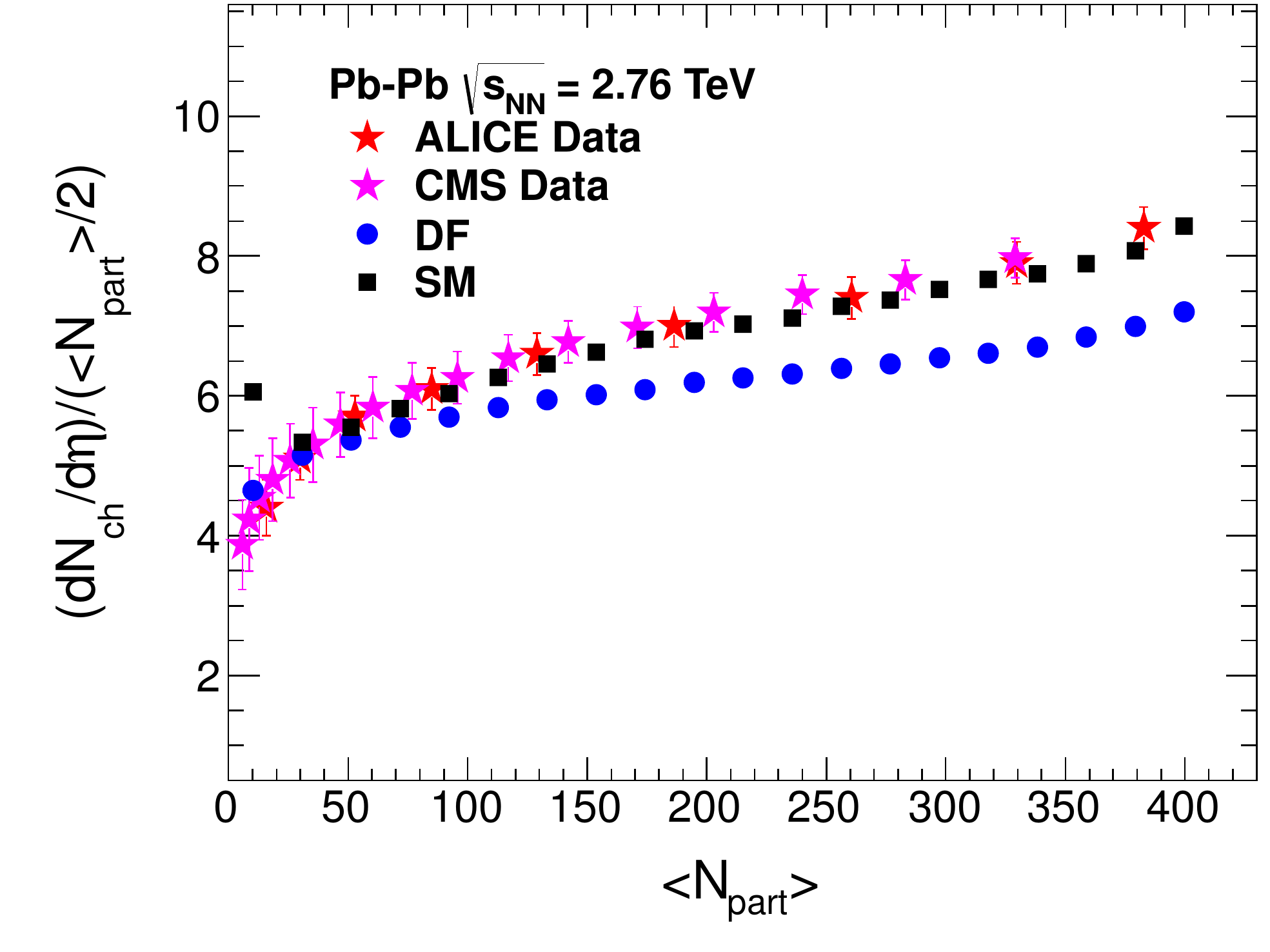}}
\caption{$dN_{ch} / d\eta$ versus $N_{part}$ plot for DF and the SM AMPT, compared with the ALICE and  CMS data}
\label{fig2}
\end{figure}

where $n_{m} \ge q$ is the bin multiplicity of the $m^{th}$  bin,  and $\langle \ldots \rangle_{h}$ is the average over all bins such that for $M^{2}$ cells
\begin{equation}
f_{q}^{e}(M)= \frac{1}{M^{2}} \sum_{m=1}^{M^{2}} n_{m}(n_{m}-1) \ldots (n_{m}-q+1)
\label{eq3}
\end{equation}
Now if $F_{q}^{e}(M)$ fluctuates from event-to-event, then the deviation of $F_{q}^{e}(M)$ from $\langle F_{q}^{e}(M) \rangle_{v}$ ($\langle \ldots \rangle_{v}$ is for averaging over all events) for each event can be  quantified using $p^{th}$ order moments of the normalised $q^{th}$ order factorial (horizontal) moments that can be defined as
\begin{equation}
C_{p,q}(M) = \langle \phi_{q}^{p}(M)\rangle_{v} 
\label{eqcpq1}
\end{equation}
where $p$ is a positive real number and 
\begin{equation}
   \phi_{q}^{p}(M) = \frac{ [F_{q}^{e}(M)]^{p}}{\langle F_{q}(M)\rangle_{v}^{p}}
   \label{eqphipq}
\end{equation}
To search for $M$-independent property of $C_{p,q}(M)$ one looks for a power-law
 behaviour of $C_{p,q}(M)$ in $M$,
\begin{equation}
  C_{p,q}(M) \propto M^{\psi_{q}(p)}
\end{equation}
this is referred to as  \textit{erraticity}~\cite{Cao:1995cf}. If $\psi_{q}(p)$ is found to have a linear dependence on $p$, then  \textit{erraticity index} $\mu_{q}$ can be defined as 
\begin{equation}
\mu_{q} =\frac{d\psi_{q}(p)}{dp} 
\label{muq}
\end{equation}
in the linear region so that it is independent of both $M$ and $p$. $\mu_{q}$ is a number that characterizes  the fluctuations of spatial patterns from evevt-to-event.  $\mu_{4}$ is observed~\cite{Hwa:2012hy} to be an effective measure to distinguish different  criticality classes, viz., critical, quasicritical, pseudocritical and non-critical, having low value for critical hadronization compared to those having random hadronization.  To a good approximation, it is observed~\cite{Hwa:2012hy} that for the model with contraction owing to confinement, $\mu_{4}$(critical and quasicritical case) = $1.87\pm0.84$  and for models without contraction $\mu_{4}$(pseudocritical and noncritical) = $4.65\pm0.06$. These model values are suggestive of the significance of erraticity index to characterize dynamical processes.
\begin{table}
\renewcommand{\arraystretch}{0.93}
\addtolength{\tabcolsep}{1.0pt}
\centering
\begin{tabular}{c c  c }
\hline  
\bf{$p_{T}$}   & \textbf{Default}         & \textbf{String Melting} \\
\textbf{window}         & $<N>$   &  $<N>$       \\
\hline
 $0.2 \le p_{T} \le 0.3$   & 285.2  & 434.8 \\
 $0.3 \le p_{T} \le 0.4$   &279.2  & 355.5 \\
 $0.4 \le p_{T} \le 0.5$   &243.7  & 271.6 \\
 $0.6 \le p_{T} \le 0.7$   &163.3  & 155.5 \\
 $0.9 \le p_{T} \le 1.0$   & 80.5  & 66.1\\
\hline
\end{tabular}
\caption{Average Multiplicity of the Simulated Data sets analyzed in different $ p_{T}$ windows}
\label{tab1} 
\end{table}
%=======================================================================
\section{Data Analysed}
 Charged particles in $|\eta| \le  0.8$ and full azimuth, generated using two versions of  A MultiPhase Transport (AMPT) model~\cite{Zhang:2000ny,Lin:2005ll}, in Pb-Pb collisions at $\sqrt{s_{NN}}=2.76$~TeV are analysed. AMPT model is a hybrid model that includes both initial partonic and the final hadronic state interactions and transition between these two phases. This model addresses the non-equilibrium many body dynamics. Depending on the way the partons hadronize there are two versions, default (DF) and the string melting (SM). In the DF version partons are recombined with their parent strings when they stop interacting and the resulting strings are converted to hadrons using Lund String Fragmentation model. Whereas in the SM version, a quark coalescence model is used to obtain hadrons from the partons. 
\begin{figure}[t]
\centerline{\includegraphics[scale=0.35]{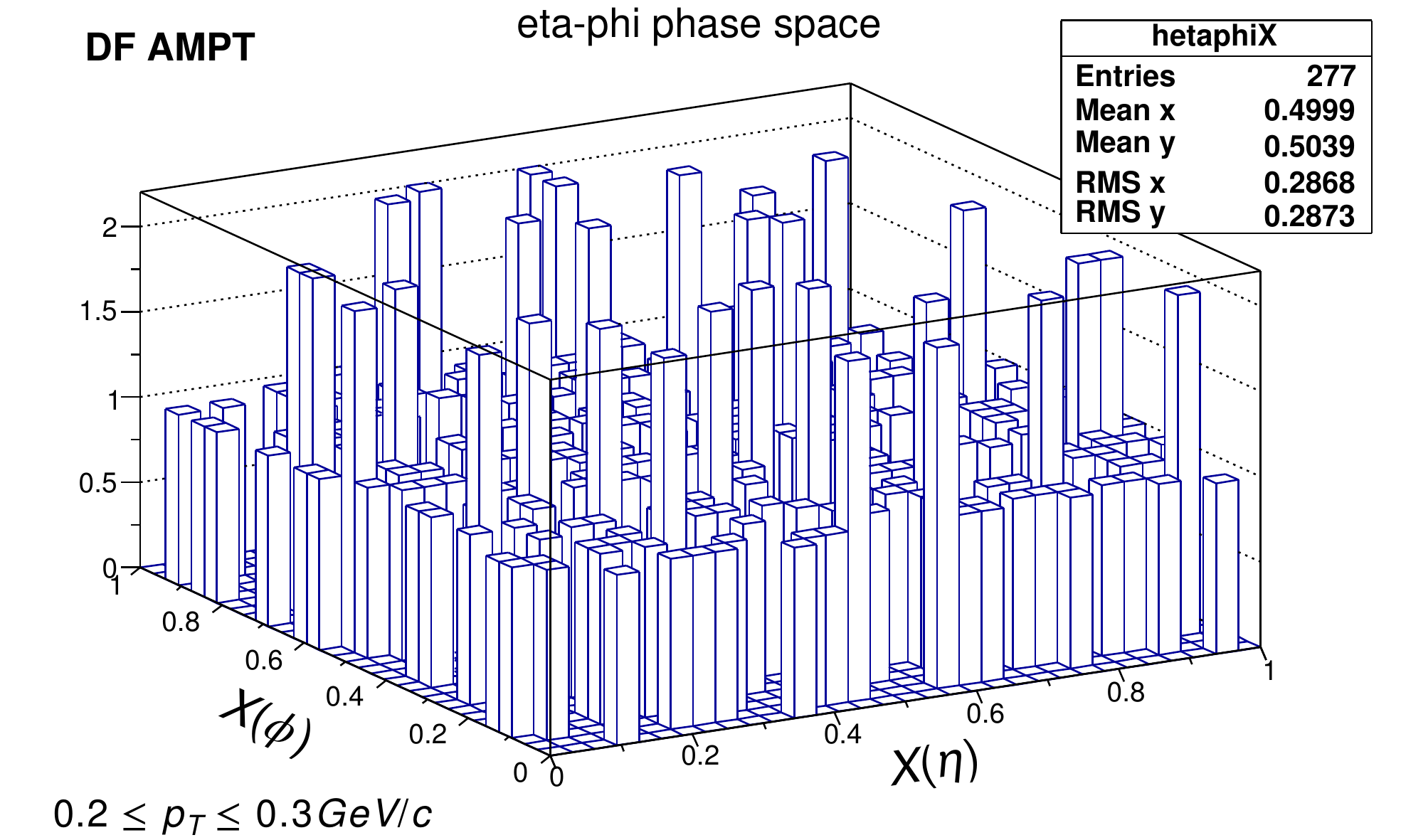}}
\caption{ ($X(\eta),X(\phi)$) phase space of an event with $M$ = 32 in DF and SM case}
\label{lego}
\end{figure}
\par
We have generated $23424$ DF and $19669$ SM events with impact parameter, $b \le 5$, using the model parameters, a = 2.2, b = 0.5, $\mu = 1.8$ and $\alpha = 0.47$ . Multiplicity, $p_{T}$, pseudorapidity and $\phi$   distributions  of the simulated events is shown in the Figure~\ref{fig1}. Charged particles generated in the $|\eta| \le 0.8$ and full azimuth having $p_{T} \le 1.0$~GeV/c in the small $p_{T}$ bins of width $0.1$~GeV/c are studied for the local multiplicity fluctuations in the spatial patterns. Five $p_{T}$  bins are considered in the present analysis, as tabulated in the Table 1, along with the average multiplicity of the generated charged particles in the respective $p_{T}$ bins. 
\par
Though AMPT does not contain the dynamics of collective interactions that are responsible for critical behaviour, but it is a good model to test the effectiveness of the methodology of analysis for finding observable signal of quark-hadron phase transition (intermittency analysis) and the quantitative measure of critical behaviour of the system (erraticity analysis) at LHC energies.

\begin{figure}
\centerline{\includegraphics[scale=0.35]{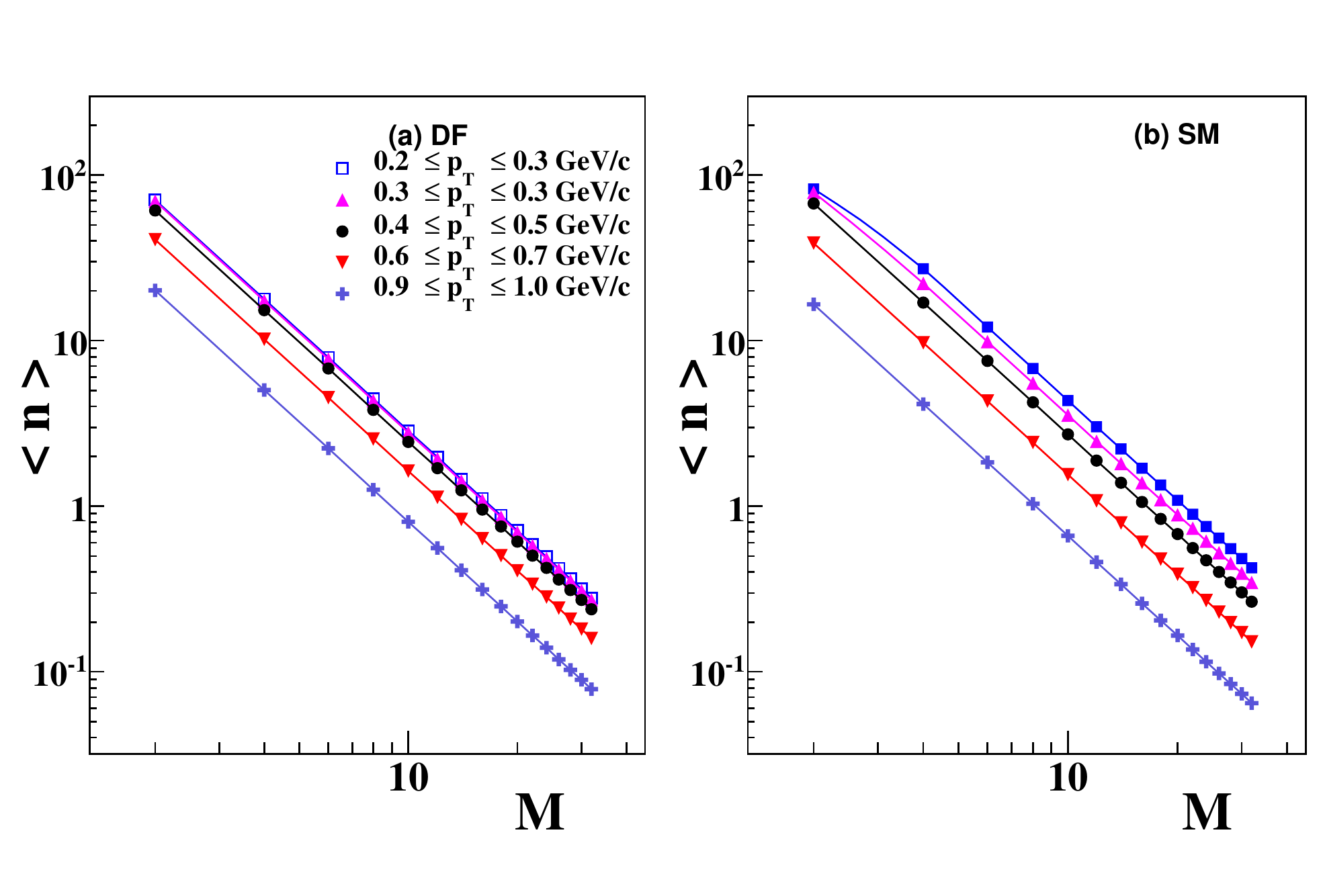}}
\caption{Average bin multiplicity dependence on $M$ for the  five $p_{T}$ bins, in case of DF and the SM AMPT model}
\label{figavbin}
\end{figure}
%===============================================================
\section{Analysis and Observations}
For an `$e^{th}$' event, the $q^{th}$ order event factorial moment ($F_{q}^{e}(M)$) as defined in Eq.\ (\ref{eq1}) are determined so as to obtain a simple characterization of the spatial patterns in two dimensional ($\eta,\phi$) space in narrow $p_{T}$ windows. However we first obtain flat single particle density distribution using cumulative variable $X(\eta)$ and $X(\phi)$~\cite{Ochs:1991zp}, which are defined as 
\begin{figure}[t]
\centerline{\includegraphics[scale=0.4]{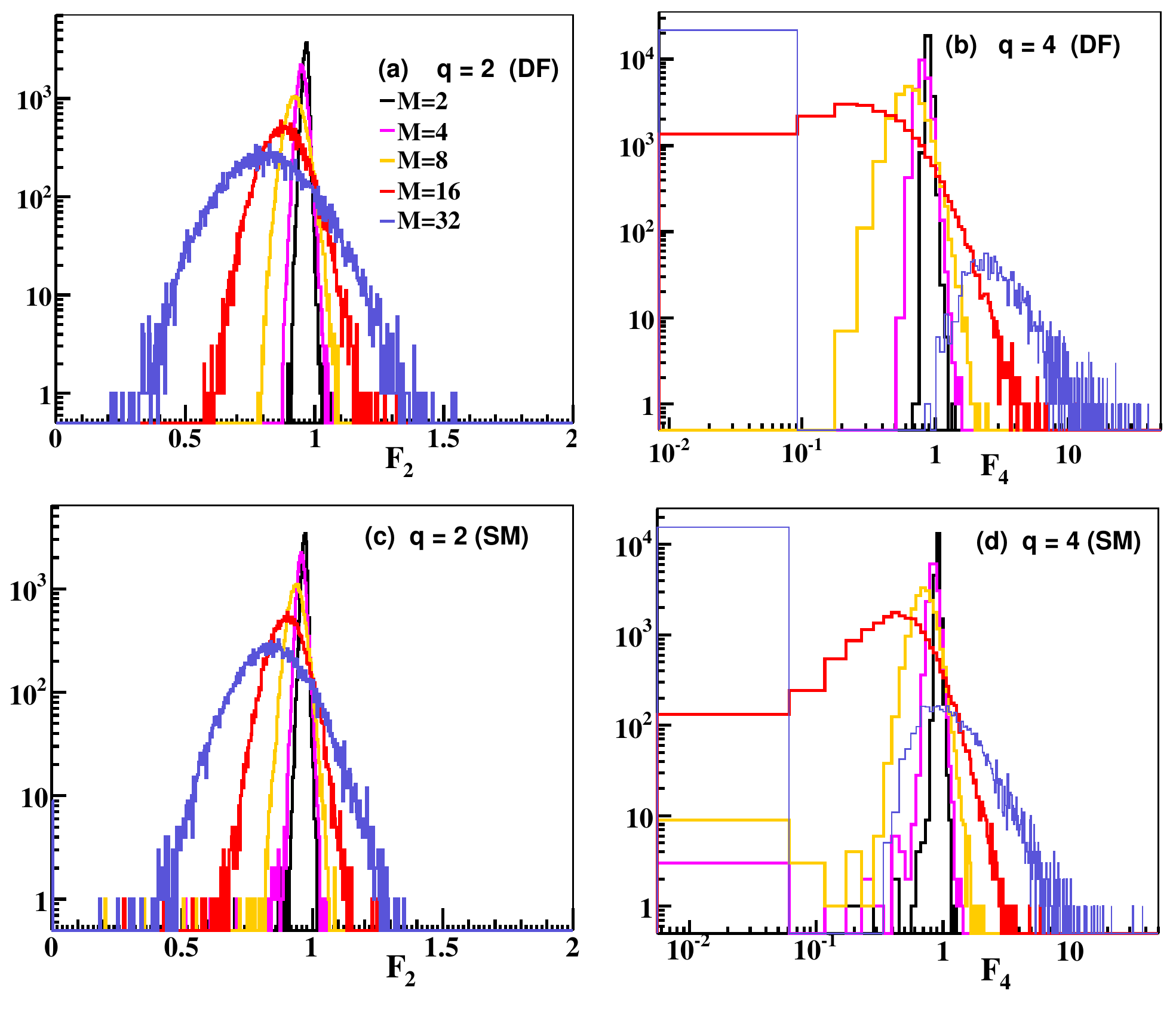}}
\caption{ $P(F_{q}^{e})$ distributions for order moment $q=2$ and $q=4$ for DF and SM ($0.3 \le p_{T} \le 0.4$).  $M$ values in multiples of 2 are shown only.}
\label{figfqe}
\end{figure}
\begin{equation}
X(y)=\frac{\int_{y_{min}}^{y}\rho(y)dy}{\int_{y_{min}}^{y_{max}}\rho(y)dy} 
\label{Xdef}
\end{equation}
here $y$ is $\eta$ or $\phi$, $y_{min}$ and $y_{max}$ denote respectively the minimum and maximum values of $y$ interval considered. $\eta$ and $\phi$ is mapped to X($\eta$) and X($\phi$) between $0$ and $1$  such that $\rho(y)$ is the single particle $\eta$ or $\phi$ density. ($X(\eta),X(\phi)$) unit square of an event in a selected $p_{T}$ window, is binned into a square matrix with $M^{2}$ bins where the maximum value that $M$ can take depends on the multiplicity in the $\Delta p_{T}$ interval and the order parameter, so that the important part of the $M$ dependence is captured. 

\par
  To give a visualization of the binning in the ($\eta,\phi$) space in narrow $p_{T}$ bin, alego plot for an arbitrary event from DF AMPT data, in  $0.2 \le p_{T} \le 0.3$  window, with M = 32,  is shown in Fig.\ \ref{lego}. 
\begin{figure}[t]
 \centerline{\includegraphics[scale=0.38]{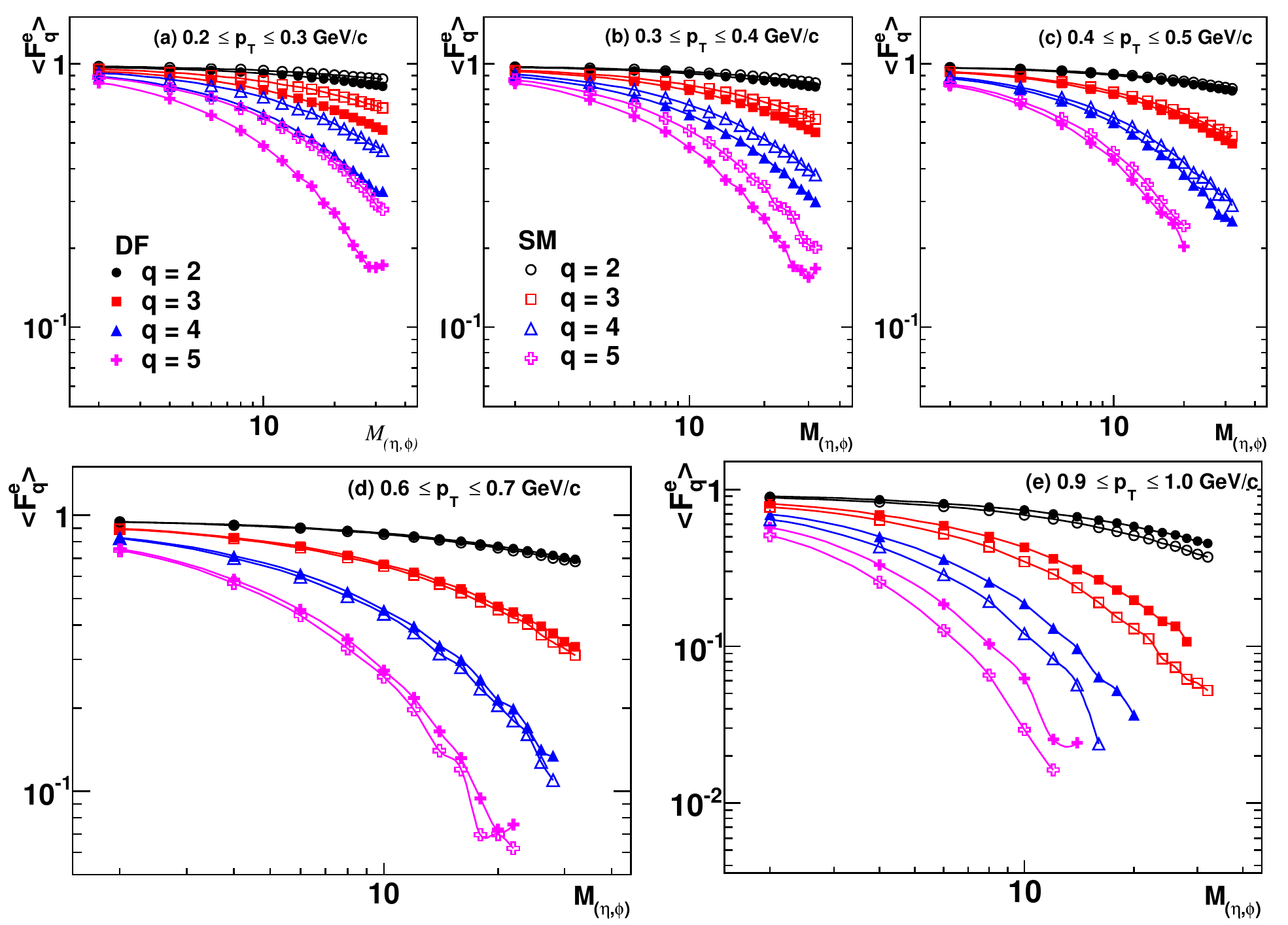}}
\caption{M Dependence of $F_{q}$, for DF as well as SM.}
\label{figfqm}
\end{figure}
\par
As value of M and $p_{T}$  increases, the $(\eta,\phi)$ space becomes empty, as is observed from Figure~\ref{figavbin} which shows the dependence of the average bin multiplicity ($\langle n \rangle$) on $M$.  Because of the denominator in Eq.\ (\ref{eq1}), a cluster of particles with multiplicity $n\ge q$ in an event would produce a large value for $F_{q}^{e}(M)$ for that event. On the other hand, if the particles are evenly distributed, $F_{q}^{e}(M)$ would be smaller. Thus the spatial pattern of the event structure should be revealed in the  distribution of $F_{q}^{e}(M)$ after collecting all events.

\par
 Study of the event factorial moment distributions, ($P(F_{q})$) reveal that  the distributions become wider as M increases and develop long tailsat higher $q$, especially at higher M values. Further the peaks of the distributions shift towards left with increase in M leading to decrease in $\langle F_{q}^{e}\rangle$ (referred to as $F_{q}$ hereafter) with M whereas the upper tails move towards right, at higher $q$ values as $M$ increases. It means that in small bins the average bin multiplicity $\left< n\right>$ is so small that when there is a spike of particles in one such bin with $n \ge q$, the non-vanishing numerator in Eq.\ (\ref{eq1}) results in a large value for $F^{e}_{q}(M)$ for that $e^{\rm th}$ event. 

\par
Dependence of $F_{q}$ on $M$ can be studied in log-log plots  as  shown in  Fig.\ \ref{figfqm} for various $p_{T}$ cuts. From the plots, it is observed that $F_{q}$'s decrease as the bin size decreases or in other words, as $M$ value increases. We observe for both the DF and SM in AMPT  that relationship between $F_{q}(M)$  and $M$ is  inverse of that in the Eq.\ (\ref{inter2}); that is
\begin{equation}
F_{q}^{\rm AMPT}(M) \propto M^{\varphi^{-}_{q}},  \qquad \varphi^{-}_{q}<0
\label{interampt}
\end{equation}
Hence, with negative $\varphi^{-}_{q}$ it is found that {\em the charged particles generated by the default and the string melting version of the AMPT model exhibit  negative intermittency}.
\par
Eq.\ (\ref{interampt}) suggests that $F_{q}^{\rm AMPT}(M)\to 0$ at large $M$ and $q$, implying that  in AMPT there are too few rare high-multiplicity spikes anywhere in phase space.  Eq.\ (\ref{interampt}) is a quantification of the phenomenon exemplified by Fig.~\ \ref{lego} for one event, and is a mathematical  characterization after averaging over many events. This same behaviour was observed in~\cite{Hwa:2012hy} for the events belonging to the non-critical class. 
\begin{figure}[t]
\centerline{\includegraphics[scale=0.4]{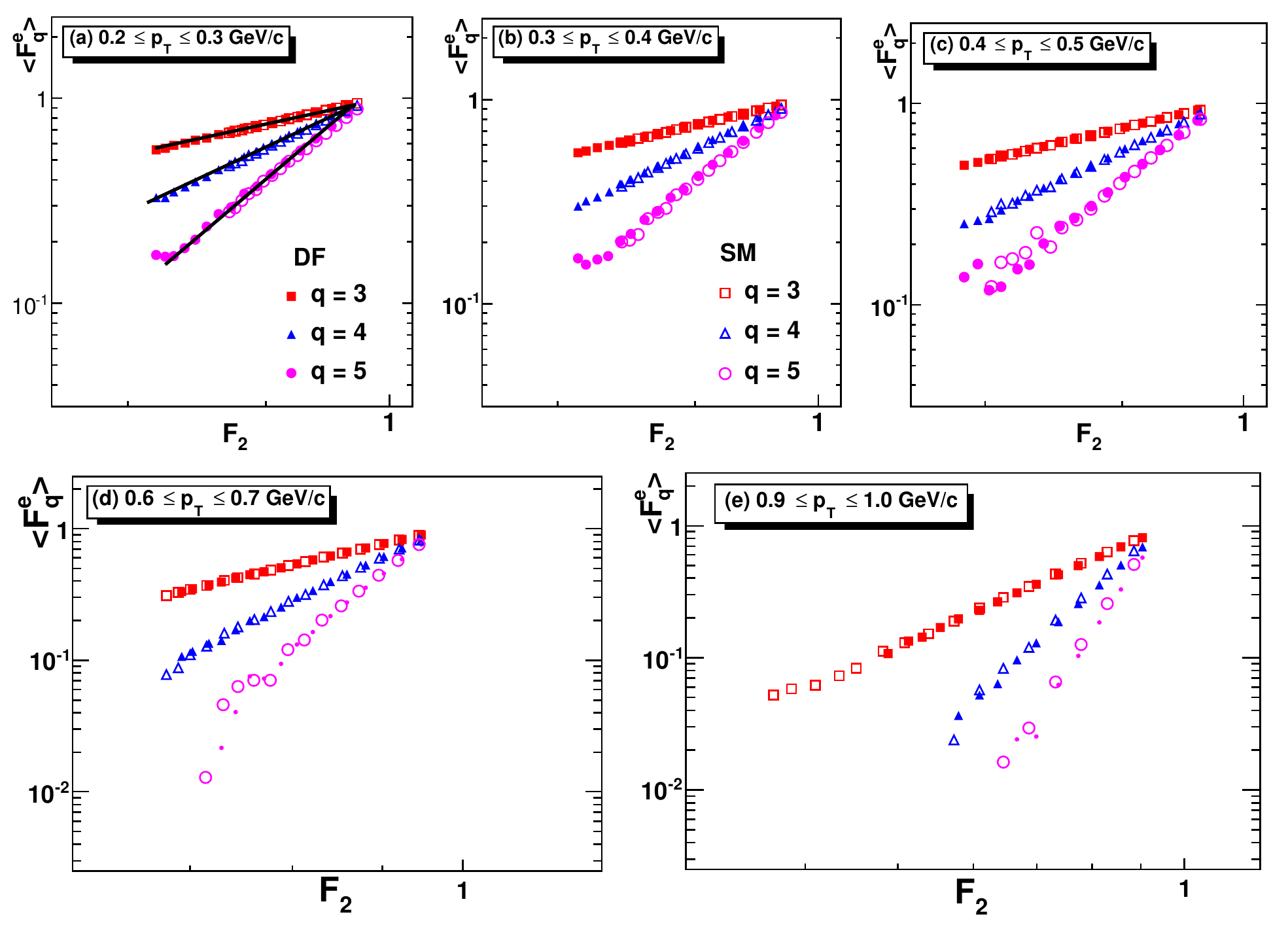}}
\caption{F-Scaling for DF as well as SM mode of AMPT}
\label{figfqf2}
\end{figure}
%=========================================
\par
\begin{figure}[t]
\centerline{\includegraphics[scale=0.38]{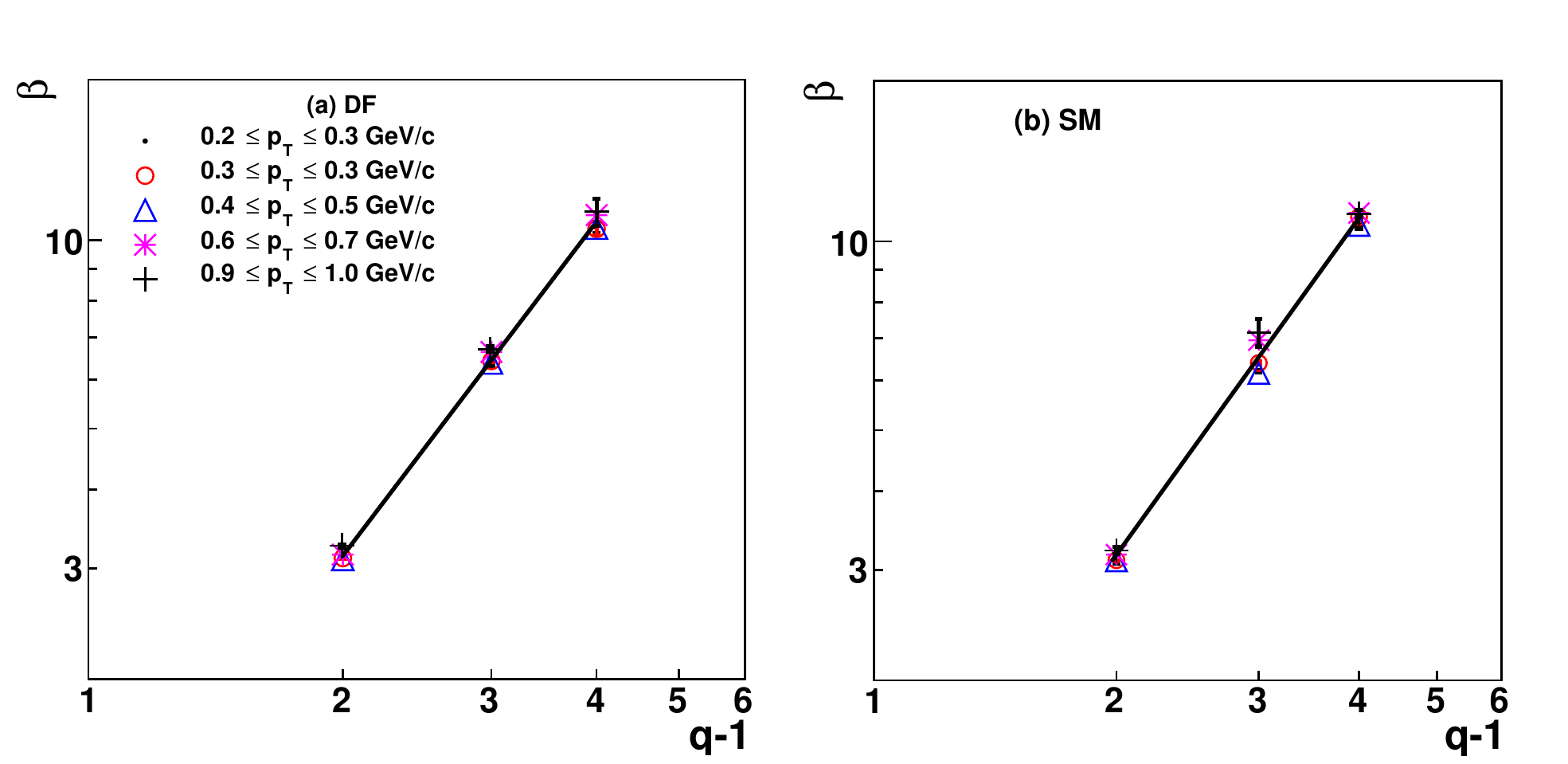}}
\caption{$\beta_{q}$ versus $(q-1)$ plot for determination of the scaling exponents.}
\label{betaqall}
\end{figure}

 We plot  $F_{q}$  versus $F_{2}$  in Fig.~\ref{figfqf2} to check F-scaling. For each set of $p_{T}$ bins linear fit has been performed to determine the value, $\beta_{q}$,  the slope, as exemplified by the straight lines in Fig.~\ref{figfqf2} (a). The dependence of $\beta_{q}$ on $(q-1)$ is shown in Fig.\ \ref{betaqall}, which exhibits good linearity in the log-log plots. Thus we obtain a scaling exponent, denoted here as $\nu_{-}$. In Table \ref{tab2} are given the value of the negative scaling exponent for different $p_{T}$ windows and  for both versions of the AMPT model studied here. Since the scaling that is there in Eq.~\ref{inter2} is different  to that we observe here for AMPT data, thus the scaling exponent obtained  here cannot be compared with the $\nu = 1.304$ for the second order phase transition, as obtained from the GL theory.
\begin{table}
\renewcommand{\arraystretch}{0.9}
\addtolength{\tabcolsep}{1.2pt}
\centering
\begin{tabular}{c c c}
\hline  
$p_{T}$ &  $\nu_{-}$             & $\nu_{-}$           \\
      &    \textbf{(Default)}  &  \textbf{(String Melting)}\\
\hline
 $0.2 \le p_{T} \le 0.3$   & $1.738 \pm 0.008$  & $1.753 \pm 0.004$  \\
 $0.3 \le p_{T} \le 0.4$   & $1.774 \pm 0.007$  & $1.793 \pm 0.005$ \\
 $0.4 \le p_{T} \le 0.5$   & $1.758 \pm 0.006$  & $1.755 \pm 0.006$  \\
 $0.6 \le p_{T} \le 0.7$   & $1.824 \pm 0.008$  & $1.869 \pm 0.016$ \\
 $0.9 \le p_{T} \le 1.0$   & $1.778 \pm 0.013$  & $1.781 \pm 0.011$  \\
\hline
\end{tabular}
\caption{Scaling exponents for negative intermittency in the Default and String Melting versions of the AMPT Model}
\label{tab2}
\end{table}
\par
Since large fluctuations result in the high $F_{q}$ tails of $P(F_{q})$, as exemplified in Fig.\ \ref{figfqe} (b) and (d), it is advantageous to put more weight on the high $F_{q}$ side in averaging over $P(F_{q})$. That is just what the double moment $C_{p,q}(M)$ does. We have determined  $C_{p,q}\rm(M)$ for q = 2, 3, 4, 5 and p = 1.0, 1.25, 1.5, 1.75 and 2.0. 
\par
\begin{figure}[t]
\centerline{\includegraphics[scale=0.48]{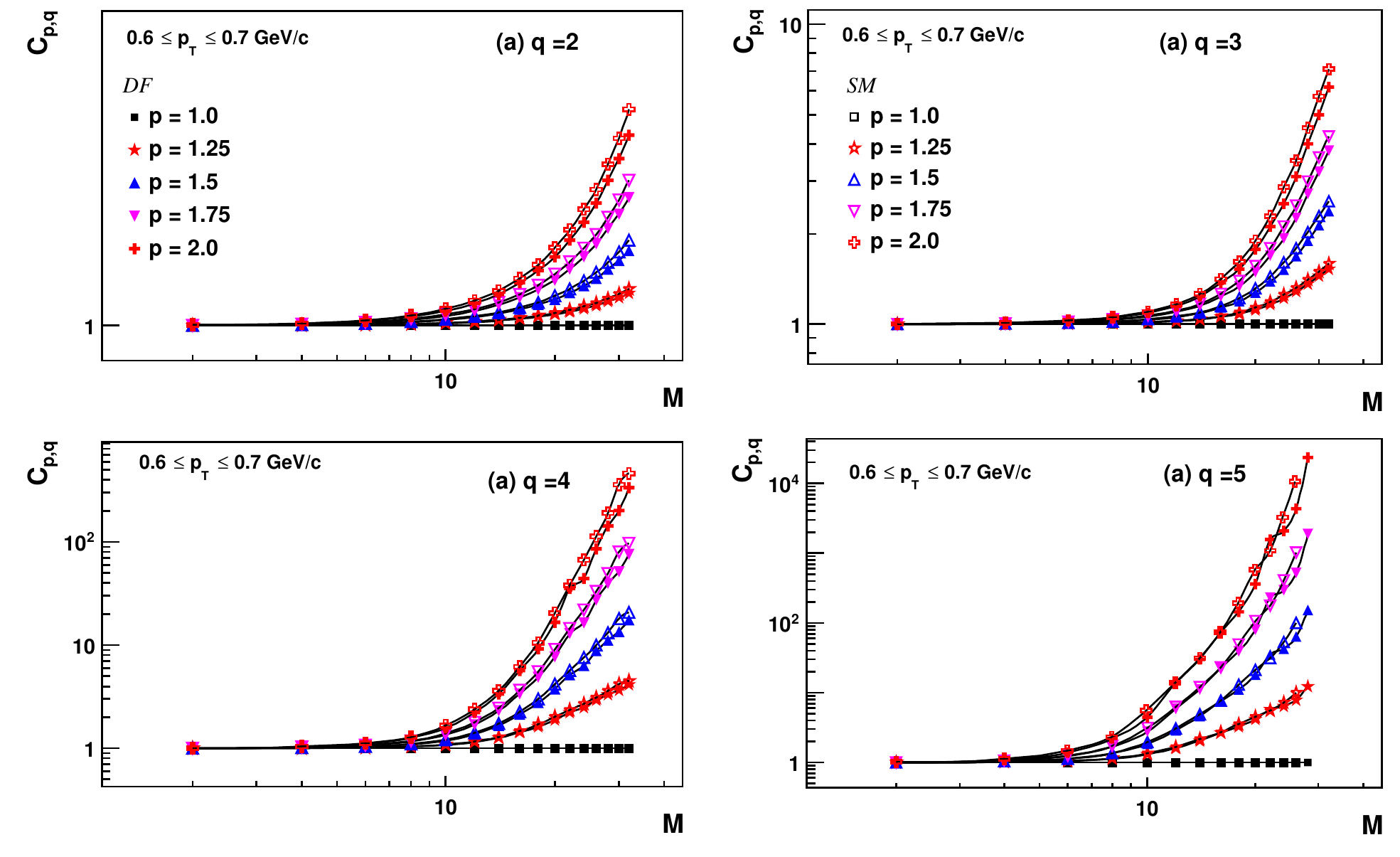}}
\caption{$M$ dependence of $C_{p,q}$ for the $p_{T}$ window $0.6 \le p_{T} \le 0.7$ GeV/c in case of DF and SM versions of  the AMPT model}
\label{cpqM6to7}
\end{figure}
\begin{figure}
\centerline{\includegraphics[scale=0.28]{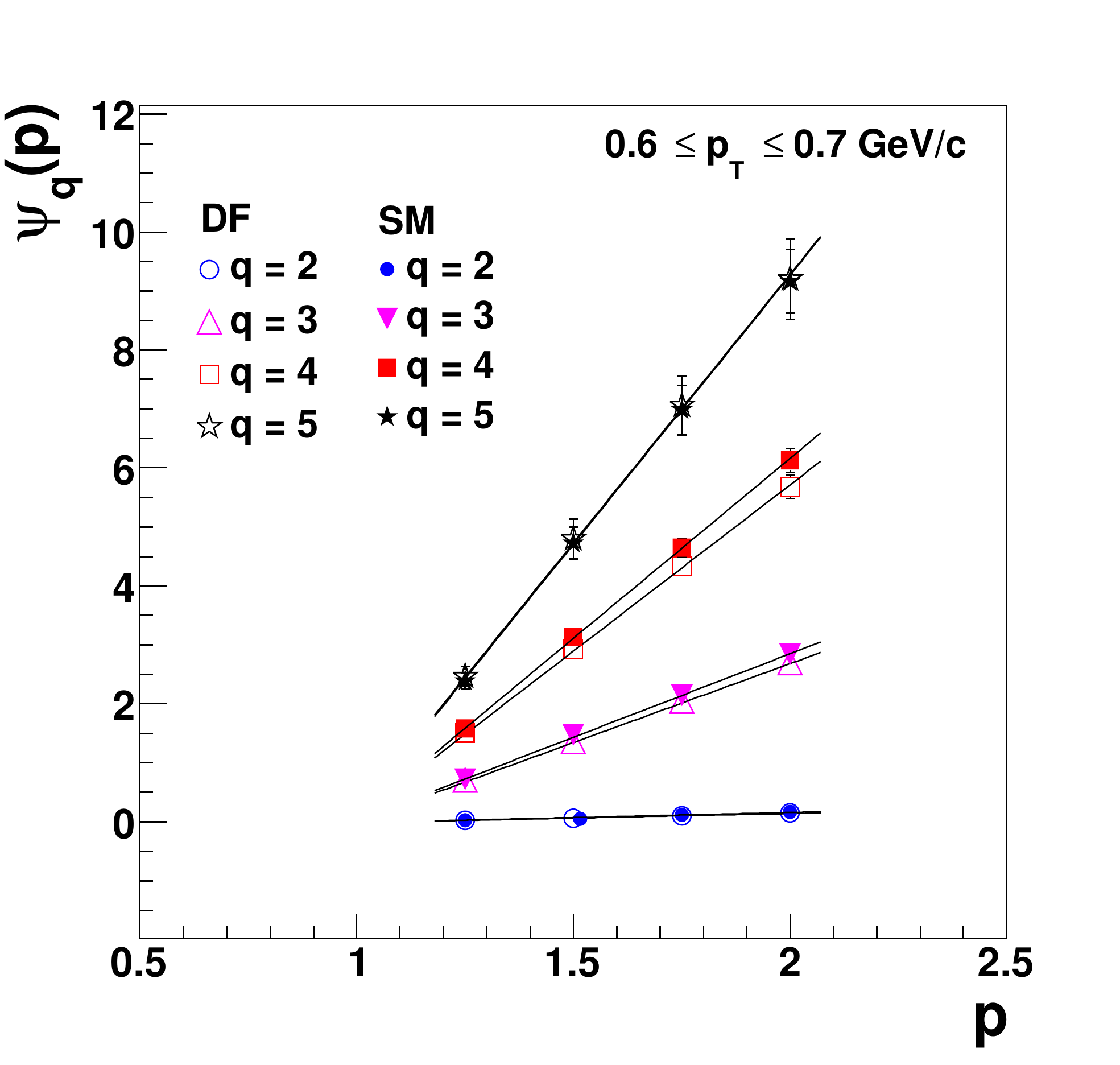}}
\caption{  The $\psi_{q}$ dependence on the $p$ for DF and SM AMPT in the $0.6 \le p_{T} \le 0.7$ GeV/c window.}
\label{psi6to7}
\end{figure}

The number of bins, $M$, takes on values from 2 to the maximum value  possible while  having reasonable  $\langle n_{m} \rangle$ such that $F_{q} \ne 0$.  To check whether $C_{p,q}\rm(M)$ follows the scaling behaviour with $M$, $C_{p,q}$ is plotted against $M$. Fig.\ \ref{cpqM6to7} (a) to (d) shows respectively, for $q$ = 2, 3, 4 and 5, the $C_{p,q}$ versus $M$ plot in the log-log scale for the window $0.6 \le p_{T} \le 0.7$ GeV/c,  and for various values of $p$ between 1 and 2.  As expected  for  all values of $q$,  for $p$ = 1.0, the $C_{p,q} = 1$. For $p > 1.0$, $C_{p,q}$ increases with $M$ and $q$ values.  Similar calculations are also done for the other $p_{T}$ windows.  In the  high $M$ region linear fits are performed  for  each $q$ and $p$ value so as to determine $\psi_{q}(p)$.
We see in Fig.\ \ref{psi6to7}  that for $0.6 \le p_{T} \le 0.7$ \ $\psi_{q}(p)$ depends on $p$ linearly for each $q$. Thus the erraticity indices defined in Eq.\ (\ref{muq}) are  determined. Similar plots are obtained for the other $p_{T}$ windows also and the values of $\mu_{q}$ are given in Table \ref{tab3}.    It can be seen from the table that as $p_{T}$ value increases, the erraticity indices increase for both versions of the AMPT model.  
\par
Comparing the $\mu_{q}$ values for the DF and SM data within the same window and for the same values of  $q$, it is observed that $\mu_{q}$ has higher values for the DF version in comparison to SM for the $p_{T}$ windows below  $ 0.6 $ GeV/c. That phenomenon is related to the average multiplicities of the two versions reversing their relative magnitudes at higher $p_{T}$. However it is to be noted from Fig.\ \ref{psi6to7} that the dependence of $\psi_{q}(p)$ on $p$ is better distinguishable for the two versions of the AMPT for only $q$ = 4. Coincidentally, as observed in~\cite{Hwa:2012hy}, $\mu_{4}$ seems to be a good measure to compare the erraticity indices of the different systems and data sets at these energies.
 %=================
%==============================

\par
We observe that the values of $\mu_{4}^{DF}$ and  $\mu_{4}^{SM}$ for all $p_{T}$ windows are larger than those obtained for the critical data set in~\cite{Hwa:2012hy}, on the same side as the non-critical case. We have found $\varphi^{-}$ to be negative because $P(F_{q})$ broadens, as $M$ increases, with  $\left< F_{q}{e}\right>$ shifting to the lower region of $F_{q}^{e}$, thus resulting in negative intermittency. We did notice that the upper tails move to the right, suggesting the presence of some degree of clustering. To emphasize that part of $P(F_{q})$ we have taken higher $p$-power moments of $\phi_{q}(M)$, which suppress the lower side of  $F_{q}$ while boosting the upper side. The scaling properties of $C_{p,q}(M)$ therefore deemphasize what leads to negative intermittency.  Thus the erraticity indices $\mu_{q}$ reveal a different aspect of the fluctuation patterns than the scaling indices $\nu^{-}$.  Our study here has revealed interesting properties of scale-invariant fluctuations that should be compared to the real data. 
\begin{table}
\renewcommand{\arraystretch}{0.8}
\addtolength{\tabcolsep}{1.2pt}
\centering
\begin{tabular}{c c c }
\hline  
                                    &      AMPT                     &    AMPT                                               \\
                                    &     {\textbf(Default)}  &  {\textbf (String Melting)}      \\
\hline
$p_{T}$           &     $\mu_{2}$               & $\mu_{2}$                     \\
\hline
$0.2 \le p_{T} \le 0.3$   & $ 0.043 \pm 0.002$  & $0.016 \pm 0.001$  \\
 $0.3 \le p_{T} \le 0.4$   & $0.045 \pm 0.002$  & $0.027 \pm 0.001$\\
 $0.4 \le p_{T} \le 0.5$   & $0.062 \pm 0.003$  & $0.048 \pm 0.002$   \\
 $0.6 \le p_{T} \le 0.7$   & $0.154 \pm 0.008$  & $0.174 \pm 0.011$   \\
 $0.9 \le p_{T} \le 1.0$   & $0.739 \pm 0.043$  & $1.014 \pm 0.064$  \\
\hline
           &     $\mu_{3}$               & $\mu_{3}$                      \\
\hline
$0.2 \le p_{T} \le 0.3$   & $ 0.901 \pm 0.081$  & $0.328 \pm 0.025$  \\
 $0.3 \le p_{T} \le 0.4$   & $0.904 \pm 0.081$  & $0.531 \pm 0.043$  \\
 $0.4 \le p_{T} \le 0.5$   & $1.304 \pm 0.118$  & $1.019 \pm 0.077$   \\
 $0.6 \le p_{T} \le 0.7$   & $2.678 \pm 0.155$  & $2.832 \pm 0.137$   \\
 $0.9 \le p_{T} \le 1.0$   & $4.502 \pm 0.147$  & $4.960 \pm 0.305$  \\
\hline
           &     $\mu_{4}$               & $\mu_{4}$                   \\
\hline
 $0.2 \le p_{T} \le 0.3$   & $4.325 \pm 0.243$  & $2.481 \pm 0.183$  \\
 $0.3 \le p_{T} \le 0.4$   & $4.532 \pm 0.234$  & $3.385 \pm 0.235$ \\
 $0.4 \le p_{T} \le 0.5$   & $5.478 \pm 0.258$  & $3.935 \pm 0.021$   \\
 $0.6 \le p_{T} \le 0.7$   & $5.640 \pm 0.203$  & $6.101 \pm 0.214$  \\
 $0.9 \le p_{T} \le 1.0$   & $7.484 \pm 0.361$  & $7.359 \pm 0.305$  \\
\hline
           &     $\mu_{5}$               & $\mu_{5}$                   \\
\hline
 $0.2 \le p_{T} \le 0.3$   & $6.202 \pm 0.302$  & $ 5.143 \pm 0.022$  \\
 $0.3 \le p_{T} \le 0.4$   & $6.150 \pm 0.175$  & $ 5.745\pm 0.312$  \\
 $0.4 \le p_{T} \le 0.5$   & $7.396 \pm 0.437$  & $ 6.159 \pm 0.280$  \\
 $0.6 \le p_{T} \le 0.7$   & $9.107 \pm 0.693$  & $ 8.360 \pm 0.533$   \\
 $0.9 \le p_{T} \le 1.0$   & $8.643 \pm 0.537$  & $ 7.655 \pm 0.358$  \\
\hline
\end{tabular}
\label{tab3}
\end{table}
\section{Summary}
Fluctuations in the spatial patterns of charged particles and their event-by-event fluctuations, as are present in the events generated using the default and string melting version of A MultiPhase Transport (AMPT) model are studied. This is a first attempt to study  intermittency and erraticity at such high energies. It is observed that as the bin size decreases, the factorial moments decrease. This behaviour is in contrast to usual properties of intermittency observed at lower energies indicating that events with localization of even moderate multiplicities in the small bins, at low $p_{T}$ are not present in the AMPT.  
Further the erraticity analysis of the model  shows that the systems generated in it is not near criticality.  The $\mu_{q}$ values determined here give the quantification of the event-by-event fluctiations in the spatial patterns of the charged particles in the midrapidity region,  which can be used effectively to compare with other models. More importantly comparison of the values with that from the LHC would help to get a better understanding of the particle production mechanism at high energies.

\end{document}